\documentclass[aps,twocolumn,prx,superscriptaddress,showpacs,longbibliography]{revtex4-1}

\usepackage{graphicx}
\usepackage{dcolumn}
\usepackage{bm}
\usepackage{upgreek}
\usepackage{paralist}
\usepackage{ifthen}

\begin{document}

\newcommand{\EECS}{\affiliation{Department of Electrical Engineering and Computer Science, Massachusetts Institute of Technology, Cambridge MA 02139}}
\newcommand{\Cornell}{\affiliation{School of Electrical and Computer Engineering, Cornell University, Ithaca NY 14853}}
\newcommand{\ElementSix}{\affiliation{Element Six, 3901 Burton Drive, Santa Clara, CA 95054, USA}}
\newcommand{\Equal}{\email{S.M and T.S. contributed equally to this work.}}

\pagestyle{empty}
\title{Scalable integration of long-lived quantum memories into a photonic circuit}

\author{Sara L. Mouradian}
\email{smouradi@mit.edu}
\EECS
\author{Tim Schr\"{o}der}
\Equal
\EECS 
\author{Carl B. Poitras}
\Cornell
\author{Luozhou Li} 
\EECS
\author{Jordan Goldstein}
\EECS
\author{Edward H. Chen}
\EECS
\author{Jaime Cardenas}
\Cornell
\author{Matthew L. Markham}
\ElementSix
\author{Daniel J. Twitchen}
\ElementSix
\author{Michal Lipson}
\email{ml292@cornell.edu}
\Cornell
\author{Dirk Englund}
\email{englund@mit.edu}
\EECS
\date{\today}

\begin{abstract}
We demonstrate a photonic circuit with integrated long-lived quantum memories. Pre-selected quantum nodes - diamond micro-waveguides containing single, stable, and negatively charged nitrogen vacancy centers - are deterministically integrated into low-loss silicon nitride waveguides. Each quantum memory node efficiently couples into the single-mode waveguide (\,$>$\,1\,Mcps collected into the waveguide) and exhibits long spin coherence times of up to 120\,$\upmu$s. Our system facilitates the assembly of multiple quantum memories into a photonic integrated circuit with near unity yield, paving the way towards scalable quantum information processing. 
\end{abstract}

\maketitle

Advanced quantum information systems, such as quantum computers~\cite{2000Bennet.NatReview} and quantum repeaters~\cite{1998Briegel_PRL_Repeater}, require multiple entangled quantum memories that can be controlled individually~\cite{NC_QCbook}. Over the past decade, there has been rapid theoretical and experimental progress in developing such entangled networks using stationary quantum bits (qubits) connected via photons~\cite{CZKM1997PRL,2008Kimble_Nat_Internet,2014Northup_NatPho_review}. Photonic integrated circuits (PICs) could provide a compact, phase-stable, and scalable architecture for such quantum networks. However, the realization of this promise requires the high-yield integration of solid state quantum memories efficiently coupled to low-loss single-mode waveguides.

A promising solid-state quantum memory with second-scale spin coherence times is the negatively charged nitrogen vacancy (NV) center in diamond~\cite{2012Maurer,2013.NComm.Bar-Gill-}. Its electronic spin state can be optically initialized, manipulated, measured~\cite{2013Doherty_ArXiv_Review}, and mapped onto nearby auxiliary nuclear memories~\cite{2007Dutt_Sci_register}. Quantum network protocols based on these unique qualities have been proposed~\cite{2005Childress_PRA_rep}, and entanglement generation and state teleportation between two spatially separated quantum nodes has been demonstrated~\cite{2014Pfaff_Sci_teleport,2013Bernien_Nat_entangle}. Translating such entanglement techniques into on-chip architectures promises scalability, but can only succeed if quantum nodes are generated with high yield. So far, yield has been inherently low due to the stochastic process of NV creation, and waveguide patterning in diamond is challenging, preventing low-loss waveguides in the optical domain around 638\,nm, the zero phonon line (ZPL) of the NV. Thus, while proof-of-principle network components have been demonstrated~\cite{2012Hausmann_NanoLett_Int,2013Hausmann_NanoLett_Ring}, the assembly of a quantum memory device into a PIC has not yet been shown. In contrast to diamond, silicon nitride (SiN)-based photonics relies on well-developed fabrication processes and is CMOS-compatible~\cite{2013Moss_NatPho_CMOSsin}. Recently, ultra-low-loss channel waveguides ($<$\,0.3\,dB/cm) and high-fidelity nonlinear devices have been demonstrated~\cite{2009Levy_NatPho_SiN}. Moreover, its large band gap ($\sim$5\,eV) and high index of refraction ($n = 2.1$) make it ideal for routing the visible emission of NVs in diamond. 


\begin{figure}
\begin{center}
\includegraphics[width = 3.33in]{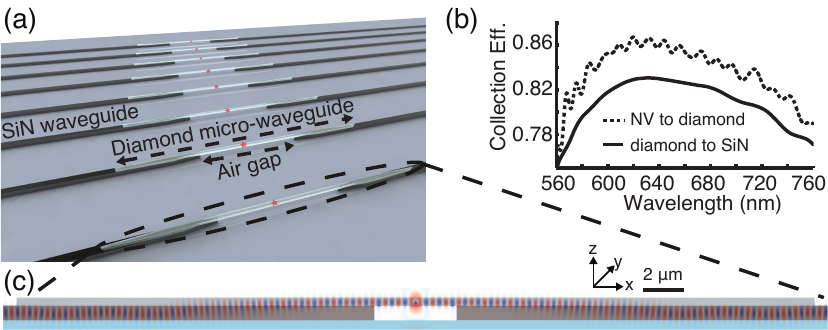}
\end{center}
\caption{(Color) (a) Sketch of a SiN PIC with multiple quantum memory nodes. (b) Simulation of the collection efficiency of NV fluorescence into the diamond $\upmu$WG (dashed) and into the SiN single-mode waveguide with optimal parameters (solid). (c) FDTD simulation ($E_x$ field) showing the mode transfer from a single mode diamond waveguide into a single mode SiN waveguide.}
\label{Fig1}
\end{figure}


Here, we address the challenges of high yield integration into low-loss photonic networks by integrating pre-selected, long-lived quantum memories based on NV centers into SiN PICs. In our approach, each quantum node consists of a diamond micro-waveguide ($\upmu$WG) suspended over a coupling region in a SiN waveguide, as illustrated in Figure \ref{Fig1}(a). The diamond $\upmu$WG supports single mode propagation over the NV emission spectrum with a cross section of 200$\times$200\,nm. Suspending it across an air gap in the SiN WG enables efficient coupling of the NV to the single optical mode in the diamond $\upmu$WG. Finite Difference Time Domain (FDTD) simulations show that with this air gap, up to 86\% (78\%) of the NV ZPL (phonon side band) fluorescence intensity at $\lambda=638$\,nm ($\lambda=$600-780\,nm) is guided into the diamond $\upmu$WG with optimal, realistic NV dipole orientation (the projection of the emission dipole  along the $\upmu$WG propagation axis is minimized for $<$100$>$ diamond $\upmu$WG) and optimal NV position (the mode maximum of the diamond $\upmu$WG), as seen in Figure \ref{Fig1}b (dashed curve). 

The SiN waveguide also supports single mode propagation over the NV emission spectrum with a cross section of 400$\times$400\,nm. Tapering of the overlapping SiN and diamond regions allows for an approximately adiabatic transition between the diamond and SiN waveguide modes. We performed FDTD simulations to optimize the tapered regions of the diamond and SiN waveguides to maximize the coupling between the diamond and SiN waveguides, as shown in Figure \ref{Fig1}. Our simulations indicate that up to 95\% of the NV fluorescence emitted into the diamond waveguide is transferred into the SiN waveguide with the optimized overlapping tapering regions shown in Figure \ref{Fig1}c (see Appendix). Thus, as shown in Figure \ref{Fig1}b (solid curve), we estimate that the total collection efficiency from the NV ZPL through the diamond $\upmu$WG into the SiN WG is 82\% for our optimized design (see Figure \ref{Fig1}c).


\begin{figure}
\begin{center}
\includegraphics[width = 3.33in]{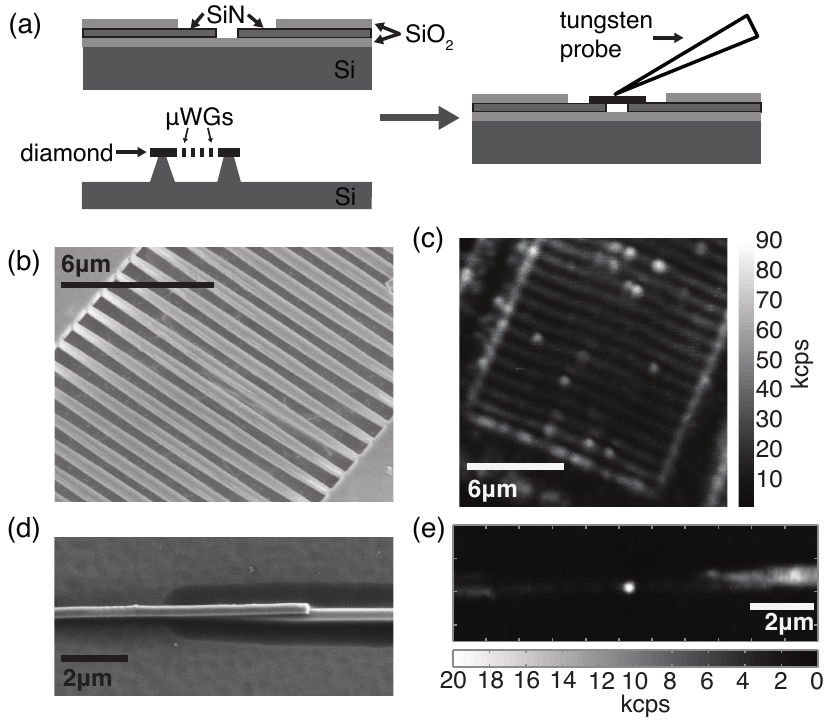}
\end{center}
\caption{(a) Illustration of quantum memory integration into a SiN PIC. Top left: A cross section view of the coupling region of a SiN PIC before integration of the diamond. Bottom left: An array of diamond $\upmu$WGs resting on an etched Si chip. Right: The integration of a diamond $\upmu$WG onto a coupling region of the SiN PIC. (b) SEM and (c) confocal PL raster scan of a diamond $\upmu$WG array. (d) SEM  and (e) confocal PL raster scan of a diamond $\upmu$WG placed over a gap in a SiN WG.}
\label{Fig2}
\end{figure}

We fabricated the diamond $\upmu$WGs from a 200\,nm-thick single crystal diamond membrane which was thinned from a  5\,$\upmu$m diamond slab produced by chemical vapor deposition. Natural NVs occurred at a concentration of approximately 0.25 NVs/$\upmu$m$^3$. A patterned silicon membrane is used as a hard etch mask~\cite{2014Li_CLEO_NV} during an oxygen plasma etch of the diamond membrane~\cite{2013.JVacSci.Li.diamond_walls} (see Appendix). We fabricated the SiN waveguides from a 400\,nm thick SiN layer deposited on silicon dioxide~\cite{2009Levy_NatPho_SiN}. The waveguides were cladded with a 3\,$\upmu$m layer of SiO$_2$ except for a 50\,$\upmu$m window over the coupling region for the integration of the diamond $\upmu$WGs. 

Figure \ref{Fig2}(b) shows a typical array of diamond $\upmu$WGs. In this experiment, we used 12\,$\upmu$m-long, 200\,nm wide $\upmu$WGs with 4\,$\upmu$m-long tapers down to 100\,nm on either side. This minimum taper size is larger than optimal, but fabrication yield was increased as the $\upmu$WGs were connected at these ends. FDTD simulations indicated that this geometry should yield a 52.5\% coupling efficiency from the NV ZPL to the SiN waveguide with optimal NV parameters (see Appendix). 

After fabrication, we characterized the $\upmu$WGs using a confocal microscope (NA = 0.9) with 532\,nm excitation. Using photoluminescence (PL) raster scans, we identified $\upmu$WGs with single NVs near the center as indicated in Figure \ref{Fig2}c. A pre-selected diamond $\upmu$WG was picked and transferred onto a SiN coupling region with a tungsten probe, where it adhered due to surface forces (see Appendix). Figure \ref{Fig2}d and e show an SEM of the transferred structure and the corresponding PL scan, respectively. This bottom-up integration process ensures that every node contains exactly one NV memory. Such high yield is not possible without pre-selection even when nitrogen atoms are spatially implanted, as the number of NVs generated remains stochastic. 


\begin{figure}
\begin{center}
\includegraphics[width =3.33in]{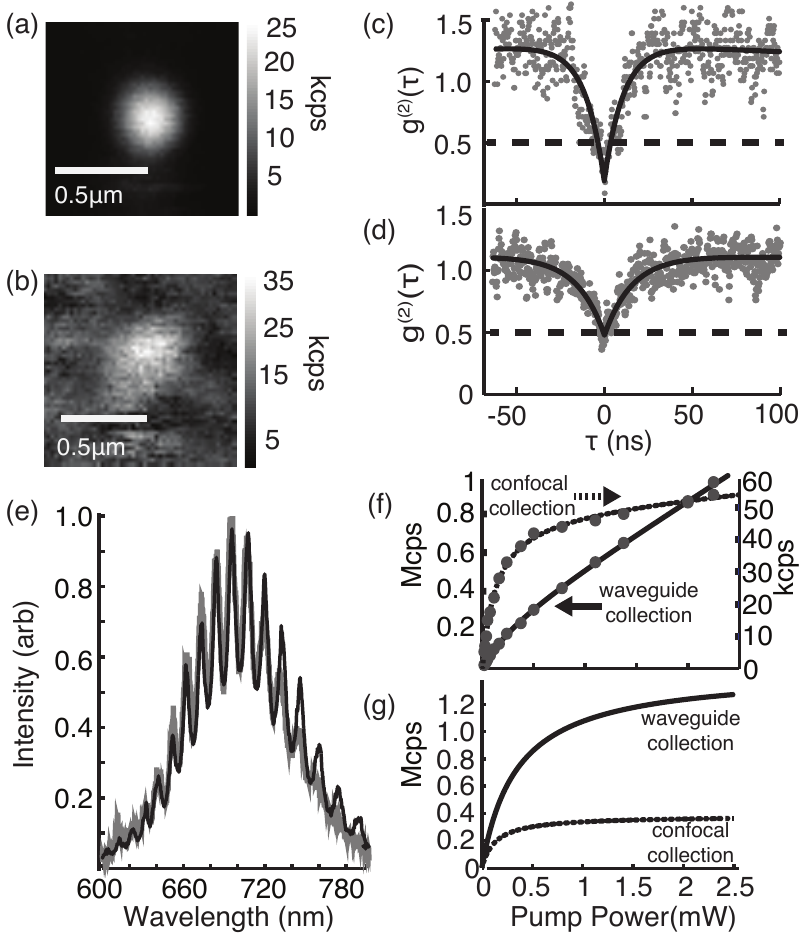}
\end{center}
\caption{(a) High resolution confocal PL raster scan of the NV with confocal collection and (b) waveguide collection. For (b), the background was subtracted using spin-selective fluorescence~\cite{2013Chen_NanoLett_Imaging,2014Sarkar_BOE_bgFreeNV}(c) normalized autocorrelation measurement g$^{(2)}(\tau)$ via confocal collection confirming single photon emission. (d) cross-correlation measurement between photons collected via the confocal setup and through the waveguide confirming that the majority of photons collected through the waveguide originate from a single NV. Solid curves in (c,d) are fits to the data. (e) Spectrum of the emitter collected via the waveguide. The solid curve is a model with parameter values taken from measured data.  (f) Saturation measurements acquired on the same emitter with confocal (dashed) and waveguide (solid) collection. (g) fits from (f) without background and corrected for measured collection losses (see text).}
\label{Fig3}
\end{figure}

Single NVs in integrated devices were excited from the top with the confocal setup described above and PL was collected via the confocal setup into a single mode fiber (confocal collection) and through the SiN waveguide into a lensed single mode fiber (waveguide collection). In both cases, the collected fluorescence was filtered with a 550\,nm long pass filter. We will focus on the optical analysis of the integrated system seen in Figure \ref{Fig2}d,e. Figure \ref{Fig3}a and b show PL raster scans of the NV under confocal excitation with confocal and waveguide collection, respectively. 

Second order correlation measurements on photons collected via the confocal setup confirm that all NVs in tested $\upmu$WGs show single emitter character, with anti-bunching as low as g$^{(2)}(0)=0.07$ (see Appendix), implying that the NVs are extremely well isolated from background sources of fluorescence. Figure \ref{Fig3}c shows the normalized auto-correlation measurement of the NV in Figure \ref{Fig3}a,b, with g$^{(2)}(0)=0.17$ at $100\,\upmu$W of 532\,nm excitation. We also performed cross-correlation measurements between photons collected via the confocal and waveguide setups. The normalized histogram in Figure \ref{Fig3}d indicates clear anti-bunching at 60$\,\upmu$W of excitation. We attribute the increase in g$^{(2)}(0)$ from the confocal auto correlation measurement to PL from the SiN waveguide due to coupling of the 532\,nm excitation laser into the WG. This PL is also observed in the spectrum collected through the waveguide (see Appendix). This background could be eliminated by fabricating a distributed Bragg reflector at the excitation laser wavelength into the SiN waveguide near the diamond coupling region~\cite{2014Harris_ArXiv,2012Wang_OptExp_DBR}. This filter could block scattered laser light from traveling through the SiN WG and exciting background fluorescence. 

Figure \ref{Fig3}e plots the spectrum of the fluorescence collected through the waveguide. An interference pattern is visible, which we attribute to an etalon effect at the diamond end facets. This etalon effect indicates an intensity reflection of $r^2 = 0.17$ and a waveguide group index of 1.7 for the 12\,$\upmu$m long diamond $\upmu$WG, which matches our expectations (see Appendix).

To evaluate the enhancement in collection efficiency through the waveguide, we performed emitter saturation measurements with both confocal and waveguide collection as seen in Figure \ref{Fig3}f. In each case, the excitation polarization was tuned to maximize the signal-to-noise ratio. For confocal collection, this entailed maximizing the NV excitation rate, while under waveguide collection this entailed limiting the coupling of the 532\,nm excitation into the waveguide to minimize background fluorescence. The optimized polarization for waveguide collection reduces the NV excitation rate, and as such increases the saturation intensity from 135\,$\upmu$W via confocal collection to 350\,$\upmu$W via waveguide collection. For confocal (waveguide) collection, 16\,kcps (55\,kcps) were detected at 60$\,\upmu$W of excitation, as used to measure the cross-correlation seen in Figure \ref{Fig3}d.

Figure \ref{Fig3}g shows the fits in Figure \ref{Fig3}f without the linear background terms, and corrected for the measured collection efficiencies of each collection pathway. The waveguide (confocal) collection pathway has a measured efficiency of 25\% (17\%). Both are measured with an Si avalanche photodiode (APD) with quantum efficiency $\eta = 0.65$. Without these system inefficiencies, we estimate that 0.38$\times10^6$ NV photons/second are collected into the objective at saturation, while 1.45$\times10^6$ NV photons/second are collected into one direction of the single mode SiN waveguide. We integrated 4 such quantum memories into the SiN PIC, each of which contained a single NV center which emitted $>1\times10^5$ photons/second into the SiN waveguide at 125\,$\upmu$W of excitation (see Appendix).


\begin{figure}
\begin{center}
\includegraphics[width = 3.3in]{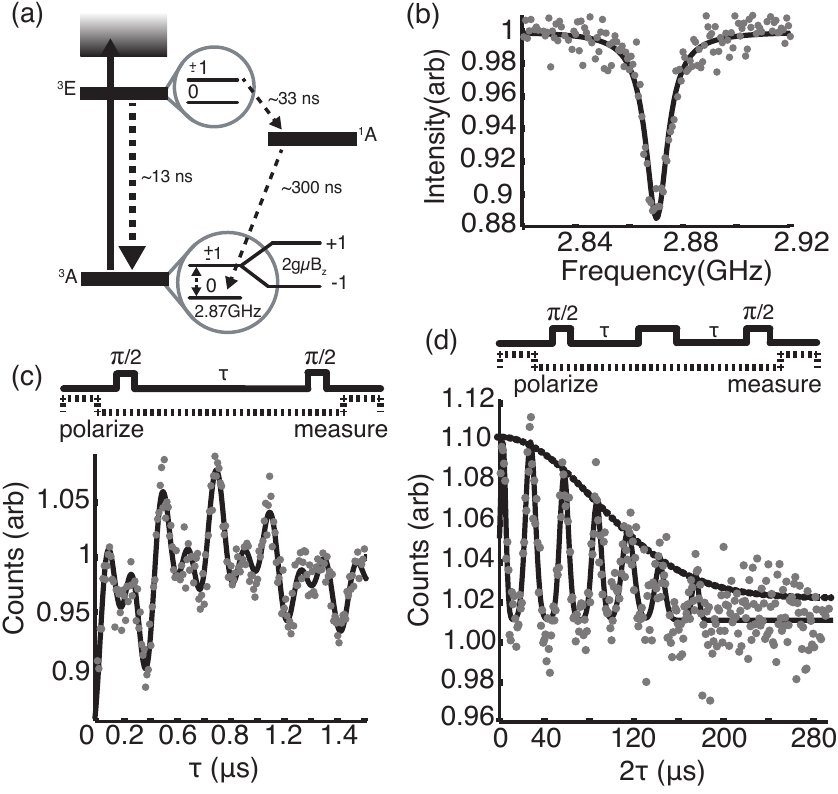}
\end{center}
\caption{(a) Level system of an NV. (b) ODMR of an NV under no magnetic field with waveguide collection. (c) The $m_s = 0$ to $1$ transition is driven off resonance and 3 Ramsey frequencies are observed due to coupling between the NV electronic spin and the host N$^{14}$ nuclear spin, with a decay due to the surrounding spin bath ($T_2^* = 2\,\upmu$s) . (d) The $\pi$ and $\pi/2$ times of an on-resonance driving field are used to construct a Hahn-Echo sequence to decouple the NV from the surrounding spin bath and measure $T_2$ $>$ 120\,$\upmu$s from the exponential decay of the coherent revivals.}
\label{Fig4}
\end{figure}

In Figure \ref{Fig4} we present the electron spin properties of the NV center in a second integrated system in which 0.8\,$\times10^6$ photons/second were collected into one direction of the waveguide at saturation. Figure \ref{Fig4}a shows the transitions of the NV center with the magnetic sublevels $m_s = -1,0,1$, obtained using optically detected magnetic resonance (ODMR)~\cite{2006Manson_PRB_levels}. Figure \ref{Fig4}b plots the ODMR fluorescence signal collected through the waveguide under continuous-wave laser excitation with no external magnetic field.

For state manipulation, the degeneracy of the $m_s = \pm 1$ levels is lifted by the application of an magnetic field of $\sim 56$\,Gauss projected onto the NV axis. A Ramsey sequence, consisting of two $\pi/2$ pulses separated by a free evolution time $\tau$, was used to probe the spin environment experienced by the NV. From this, we deduce an ensemble phase coherence time $T_2^* \simeq 2\,\upmu$s. We also performed Hahn-echo measurements~\cite{2006Hanson_PRB_echo}, which indicate a spin coherence time of $T_2 \simeq 120\,\upmu$s, see Figure \ref{Fig4}d. This long spin coherence time is similar to values observed in the parent diamond crystal~\cite{2012.NJP.diamond_slab}. We anticipate that using isotropically purified $^{12}$C diamond, together with dynamical decoupling, should enable spin coherence times in excess of tens of milliseconds~\cite{2009Balasubramanian}.

In conclusion, we have realized the integration of multiple quantum nodes into a photonic network. Each integrated node contains a single high quality solid-state qubit. The design of the diamond-SiN interface allows for efficient coupling of photons emitted by the NV center into the single mode SiN waveguide, which itself exhibits low propagation loss ($\sim0.3$\,dB/cm) and enables high coupling to single mode fiber ($\sim3$\,dB loss). This high collection efficiency into a single spatial mode indicates the promise of this system as an efficient source of single photons in a single spatial mode. Moreover, our experimental results show that each integrated node contained one long-lived quantum memories. Finally, this method can be generalized to other systems (e.g. photonic crystal cavities containing single emitters) to integrate pre-screened functional nodes into high quality PICs with essentially unity yield, paving the way towards scalable on-chip quantum networks.

\begin{acknowledgments}
S.M. was supported by the AFOSR Quantum Memories MURI. T.S. was supported by the Alexander von Humboldt Foundation. E.H.C. was supported by the NASA Office of the Chief Technologist's Space Technology Research Fellowship. Fabrication and experiments were supported in part by the Air Force Office of Scientific Research (AFOSR Grant No. FA9550-11-1-0014, supervised by Gernot Pomrenke). Research carried out in part at the Center for Functional Nanomaterials, Brookhaven National Laboratory, which is supported by the U.S. Department of Energy, Office of Basic Energy Sciences, under Contract No. DE-AC02-98CH10886. Research also carried out in part at the Cornell NanoScale Facility, a member of the National Nanotechnology Infrastructure Network, which is supported by the National Science Foundation (Grant ECCS-0335765). MLM and DJT was supported by the DARPA Quiness program. We thank Matt Trusheim for valuable discussions.
\end{acknowledgments}

\section*{Appendix}
\appendix
\section{Simulations}

To predict and optimize the coupling efficiency from an NV center to a SiN WG, we simulate our systems using FDTD computations. The NV center is represented as an electric dipole placed in the center of the of a 200\,nm x 200\,nm diamond $\upmu$WG ($n=2.4$), oriented perpendicular to the propagation axis of the $\upmu$WG and 35$^{\circ}$ off-horizontal. This is consistent with a diamond slab oriented in the $<$100$>$ direction, as we use in our experiment. We placed Poynting flux monitors (i) to either side of the NV, overlapping the diamond $\upmu$WG before the SiN WG begins, (ii) at each end of the SiN WG,  and (iii) surrounding the entire structure. These are used to monitor where electro magnetic power is lost; the ratio of (i) to (iii) gives the NV coupling efficiency to the diamond $\upmu$WG, and the ratio of (ii) to (iii) yields the total coupling efficiency of the device.

\begin{figure}
\begin{center}
\includegraphics[width = 2.5in]{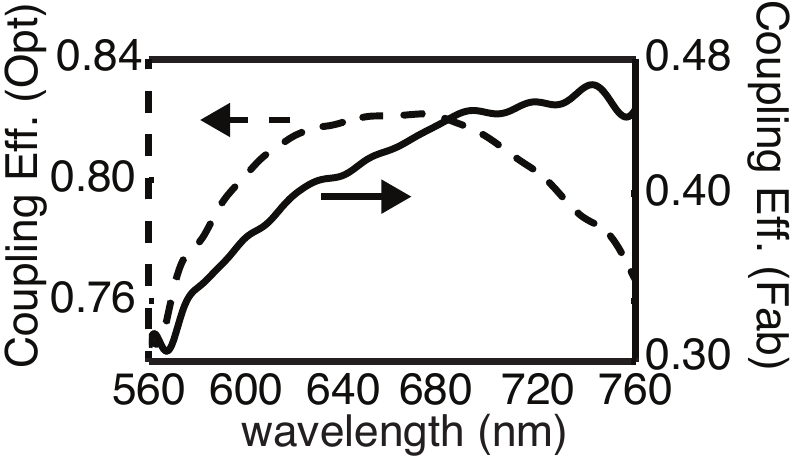}
\end{center}
\caption{Simulated coupling efficiencies from the diamond into the SiN of the optimized (dashed) and fabricated (solid) devices.}
\label{Sims}
\end{figure}

The optimized device geometry was determined by evaluating the coupling efficiency from the fundamental TE mode of the diamond $\upmu$WG to the SiN WG while sweeping the diamond $\upmu$WG and SiN WG taper lengths. The tapers are aligned such that the taper regions in SiN and diamond do not overlap, minimizing the photon loss. Based on these results, we chose a $\upmu$WG  taper length of 6\,$\upmu$um and a SiN WG taper length of 5\,$\upmu$m. Finally, we use a 2\,$\upmu$m gap in the WG to maintain high coupling from the NV into the diamond $\upmu$WG. This results in a $\upmu$WG which is 24\,$\upmu$m long, and yields an overall dipole-to-WG coupling efficiency of 83\% (41.5\% on each side), as seen in Figure \ref{Sims}. We also simulated the structure used in the experiments reported here, which was shorter, and had blunt tips (tapered down to 100\,nm instead of 0\,nm), increasing losses. This gives a coupling efficiency from the diamond $\upmu$WG to the SiN WG of 52.5\%, as seen in Figure \ref{Sims}.

\section{Fabrication}
We begin fabrication with a 5\,$\upmu$m ultra-pure diamond slab produced by chemical vapor deposition. The diamond slab is polished with Ar and Cl$_2$ down to a final thickness of 200\,nm. This membrane is patterned with oxygen plasma, using a transferrable patterned silicon membrane as an etch mask, a technique that is introduced and explained in previous publications~\cite{2014Li_CLEO_NV}

\section{Diamond Integration}
The diamond $\upmu$WGs are detached and picked up from the initial array with a tungsten microprope (Ted Pella) mounted to a 3-axis piezo micromanipulator and rotation stage. The $\upmu$WG is transferred to a SiN chip, which sits on a 2-axis and rotation stage. The diamond $\upmu$WG is aligned to the coupling region of a SiN waveguide and placed in its center. Adhesion is promoted via plasma cleaning of the SiN prior to placement.

\section{Photon correlation measurements}
\begin{figure}
\begin{center}
\includegraphics[width = 2.5in]{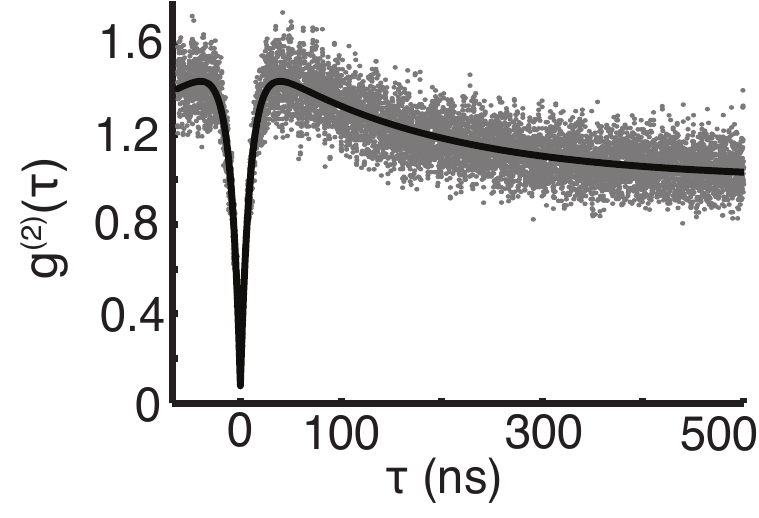}
\end{center}
\caption{Autocorrelation measurement via confocal collection of an NV in a diamond $\upmu$WG showing superb single photon character with anti-bunching down to 0.07.}
\label{G2}
\end{figure}

Photon correlation measurements were performed with a Hanbury Brown and Twiss setup. In auto-correlation measurements, the PL collected via the confocal setup was coupled to a fiber beam splitter, and the two output arms were coupled to avalanche photon detector (APD) modules (Perkin Elmer). In cross-correlation measurements between confocal and waveguide collection channels, the two single mode collection fibers are directly coupled into separate APDs. A histogram of arrival times was assembled using a time-correlated counting module in start-stop mode. Figure \ref{G2} shows an NV in an integrated nanowire exhibiting excellent single photon character with an anti-bunching dip down to 0.07. 

\section{Waveguide spectra}
\begin{figure}
\begin{center}
\includegraphics[width = 2.5in]{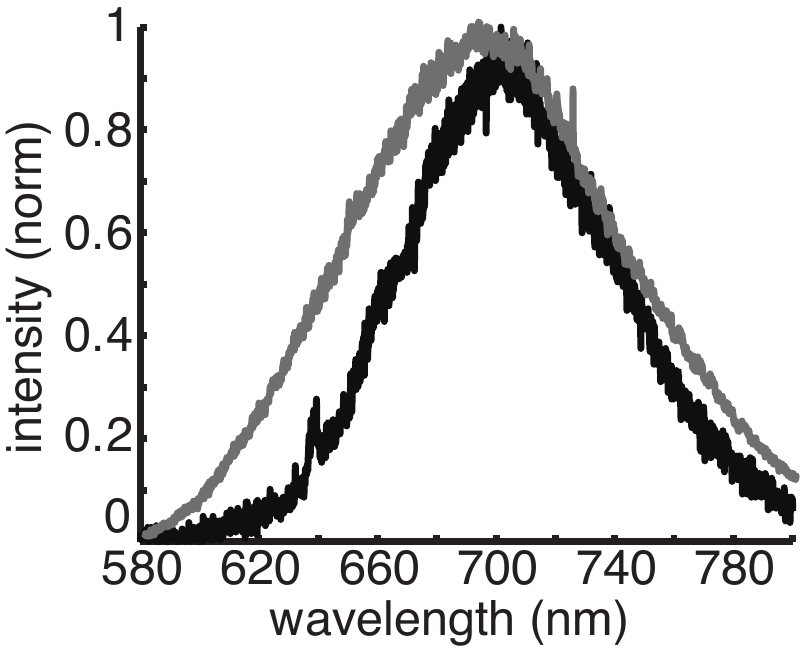}
\end{center}
\caption{Normalized spectra of (a) the NV via confocal collection (black) and (b) the SiN waveguide fluorescence (grey) under 532\,nm excitation.}
\label{Spec}
\end{figure}

The PL collected through the waveguide consists of PL originating from the diamond $\upmu$WG and PL caused by laser propagation through the SiN waveguide. The PL originating from the diamond $\upmu$WG experiences an etalon effect due to the diamond $\upmu$WG ends. The detected spectrum is modeled as 
\begin{equation}
\text{WG}(\lambda_0) = a\left| \frac{(1-r)^2 e^{-2i\pi n L/ \lambda_0}}{1-r^2 e^{-4i \pi n L / \lambda_0}}\right|^2 \text{NV}(\lambda_0)+b\text{BG}(\lambda_0)
\end{equation}
where NV$(\lambda_0)$ is the spectrum collected via the confocal setup (Figure \ref{Spec}a) and BG$(\lambda_0)$ is the spectrum of the SiN fluorescence collected through the waveguide (Figure \ref{Spec}b). $n = 1.71$ is the expected effective refractive index of the diamond waveguide mode determined via eigenmode analysis of the waveguide, and $L=12\,\upmu$m is the total length of the diamond waveguide. 

\section{Scalability}
\begin{figure}
\begin{center}
\includegraphics[width = 2.5in]{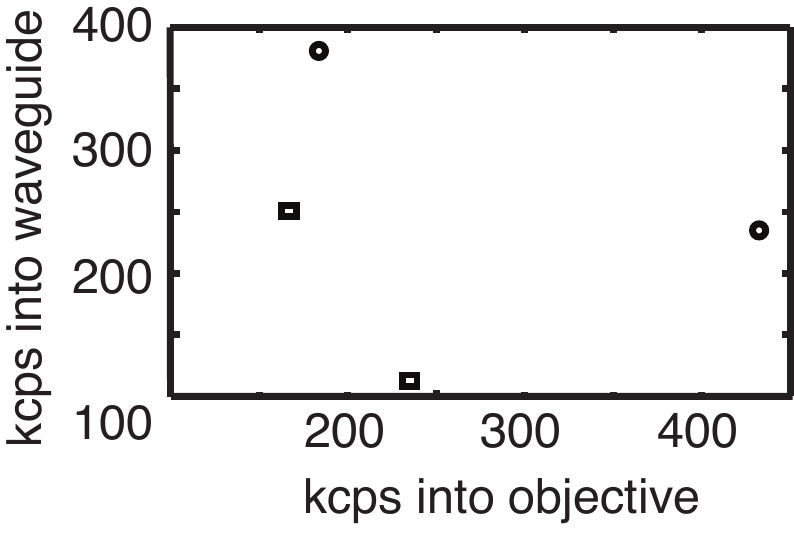}
\end{center}
\caption{Counts collected into a free space objective vs. counts collected into the single mode SiN waveguide for 4 integrated NV centers at 125\,$\upmu$W of excitation. Circles are calculated from saturation curves, where as squares are measured at 125\,$\upmu$W.}
\label{Scale}
\end{figure}
To show the scalability of our method, we integrated 4\,$\upmu$WGs into the same SiN PhC and measured NV properties of each through confocal and waveguide collection. Full saturation curves were measured for the points marked with circles. We subtracted the background, and accounted for measured system losses to arrive at the value for counts collected at 125\,$\upmu$W excitation. Auto and cross-correlation measurements were taken for the two points marked with squares at 125\,$\upmu$W excitation. The depth of the anti-bunching allowed us to subtract background, and measured system losses were accounted for to find the photons per second collected into the objective, or into the SiN waveguide. All showed efficient coupling into the SiN waveguide, with differences perhaps due to NV positioning and orientation.

\bibliography{SiN_NV_rev_v1,revtex-custm}

\end{document}